# Magnetopause expansions for quasi-radial interplanetary magnetic field: THEMIS and Geotail observations

A. V. Suvorova[1,2], J.-H. Shue[1], A. V. Dmitriev[1,2], D. G. Sibeck[3], J. P. McFadden[4], H. Hasegawa[5], K. Ackerson[6], K. Jelínek[7], J. Šafránková[7], Z. Němeček[7]

[1]*Institute of Space Science, National Central University, Jhongli, Taiwan*
[2]*Skobeltsyn Institute of Nuclear Physics, Moscow State University, Moscow, Russia*
[3]*NASA Goddard Space Flight Center, Greenbelt, Maryland, USA.*
[4]*Space Sciences Laboratory, University of California, Berkeley, California, USA*
[5]*Institute of Space and Astronautical Science, JAXA, Sagamihara, Japan.*
[6]*Department of Physics and Astronomy, University of Iowa, Iowa City, Iowa, USA*
[7]*Faculty of Mathematics and Physics, Charles University, Prague, Czech Republic*

Short title : MAGNETOPAUSE EXPANSIONS

**Abstract** We report THEMIS and Geotail observations of prolonged magnetopause (MP) expansions during long-lasting intervals of quasi-radial interplanetary magnetic field (IMF) and nearly constant solar wind dynamic pressure. The expansions were global: the magnetopause was located more than 3 $R_E$ and ~7 $R_E$ outside its nominal dayside and magnetotail locations, respectively. The expanded states persisted several hours, just as long as the quasi-radial IMF conditions, indicating steady-state situations. For an observed solar wind pressure of ~1.1-1.3 nPa, the new equilibrium subsolar MP position lay at ~14.5 $R_E$, far beyond its expected location. The equilibrium position was affected by geomagnetic activity. The magnetopause expansions result from significant decreases in the total pressure of the high-$\beta$ magnetosheath, which we term the low-pressure magnetosheath (LPM) mode. A prominent LPM mode was observed for upstream conditions characterized by IMF cone angles less than 20 ~ 25°, high Mach numbers and proton plasma $\beta \lesssim 1.3$. The minimum value for the total pressure observed by THEMIS in the magnetosheath adjacent to the magnetopause was 0.16 nPa and the fraction of the solar wind pressure applied to the magnetopause was therefore 0.2, extremely small. The equilibrium location of the magnetopause was modulated by a nearly continuous wavy motion over a wide range of time and space scales.

*Keywords:* magnetopause, magnetosheath, radial interplanetary magnetic field

## 1. Introduction

Global expansions of the magnetopause (MP), formed in response to the interaction between the solar wind (SW) and the Earth's magnetosphere, are mainly associated with low dynamic pressures (<1 nPa) in tenuous solar wind flows [*Richardson et al.*, 2000; *Terasawa et al.*, 2000; *Lockwood*, 2001]. This fundamental interaction mode can be quasi-steady when SW conditions are nearly constant for a long time (about one hour and more). However, *Fairfield et al.* [1990] indicated that radial IMF orientations can also cause MP expansions. They have shown that pressure/density perturbations produced in the subsolar foreshock correlate with dayside magnetospheric magnetic field variations. They infer that the foreshock pressure fluctuations convect through the subsolar bow shock into the magnetosheath and impinge on the subsolar magnetosphere. Other studies, showed that this interaction mode is often unsteady, resulting in multiple MP crossings with interarrival times on the order of a few minutes [*Fairfield et al.*, 1990; *Sibeck*, 1995; *Russell et al.*, 1997, *Němeček et al.*, 1998].

The location of foreshock upstream from the bow shock is controlled by the angle $\theta_{Bn}$ between the IMF and the local normal to the bow shock. In the subsolar region this angle coincides with the cone angle between the IMF vector and the Earth-Sun line. When the angle $\theta_{Bn}$ is small, the local bow shock is quasi-parallel ($Q_{\parallel}$). When the IMF is radial (aligned with the Sun-Earth line), the $Q_{\parallel}$ foreshock forms upstream of the subsolar bow shock. The $Q_{\parallel}$ foreshock exhibits strong wave activity that is swept downstream into the magnetosheath turbulence, but contrast to the much quieter quasi-perpendicular ($Q_{\perp}$) shock for IMF orientations perpendicular to the local bow shock normal [*Wilkinson*, 2003]. *Fairfield et al.* [1990] demonstrated that plasma densities and dynamic pressures diminish within the foreshock, and suggested that this might result in a decrease in the subsolar magnetosheath pressure. If so, the IMF orientation may control the pressure applied to the dayside magnetosphere. According to this hypothesis, during radial (transverse) IMF the magnetosheath pressure applied to the magnetopause should be smaller (higher). Careful study of magnetopause positions as a function of IMF cone angle can verify this hypothesis.

Comprehensive studies of foreshock effects on the magnetosheath, magnetopause, and magnetosphere were presented by numerous authors [*Sibeck et al.*, 1989; *Sibeck*, 1992 1995; *Sibeck and Gosling*, 1996; *Russell et al.*, 1997; *Němeček et al.*, 1998; *Zastenker et al.*, 1999, 2002; *Shevyrev and Zastenker*, 2005; *Shevyrev et al.*, 2007]. The dayside MP moves in response to variations of the IMF cone angle [*Sibeck*, 1995]. MP motion attains greater amplitudes behind the foreshock, where amplitude vary from 0.2 to 0.8 $R_E$ [*Russell et al.*, 1997]. *Laakso et al.* [1998] and *Merka et al.* [2003] reported examples of even larger-amplitude (~2 $R_E$) MP oscillations for quasi-radial IMF orientations. Based on indirect estimates, *Merka et al.* [2003] proposed a bullet-like shape for the expanded magnetopause. They assumed that the unusual MP motion was related to a decrease of the magnetosheath pressure behind the $Q_{\parallel}$-shock. This assumption followed the ideas of *Fairfield et al.* [1990]. However, there were no magnetosheath data, which could confirm or disprove that assumption.

These results lead one to expect depressed total pressures within the magnetosheath during intervals when the IMF has a radial orientation. Two important questions remain open: (1) What fraction of the solar wind dynamic pressure is applied to the magnetosphere by the magnetosheath during intervals of radial IMF orientation? and (2) What is the average location of the magnetopause under these conditions? These effects are absent from global MHD codes and have not yet been addressed by global kinetic or hybrid codes. MP dynamics and the properties of the adjacent magnetosheath for radial IMF conditions remain almost unexplored mainly due to insufficient experimental data in the subsolar region.

The THEMIS mission [*Angelopoulos*, 2008] provides a large database of observations for detailed studies of the MP and magnetosheath. Using THEMIS and Geotail, we investigate three cases of very large MP expansion occurring for prolonged quasi-radial IMF and nearly steady SW dynamic pressures. We demonstrate that the bow shock and magnetopause lie far beyond of their expected positions. The MP expansions are found to be quasi-steady and long-lasting. We show and quantify dramatic decreases in the magnetosheath total pressure induced by rotations to quasi-radial IMF orientations rather than by decreases in the SW dynamic pressure.

## 2. Experimental data

We analyze three events on 16 July, 4 and 8 August 2007, which are accompanied by long-lasting (up to a few hours) quasi-radial IMF orientations (the cone angle is less than 30°). Solar wind and geomagnetic conditions for these time intervals are presented in Figures 1-3. During these intervals, ACE was located at GSM (225, -2, 23), (227, 28, 4), and (226, 23, 13) $R_E$, respectively, while Wind was located at (253, -67, 16), (228, -95, 33), and (232, -97, 13) $R_E$, respectively. Comparing the Wind and ACE data, we find that averaged values for SW dynamic pressure agree to within ~20%, although the two spacecraft often observe different transient variations in the plasma parameters. The IMF demonstrates higher variability and larger differences. Though, the clock and cone angles measured by ACE and Wind coincide well within some intervals. The observed differences in SW plasma and IMF parameters are due to very large distance between the monitors [e.g. *Richardson and Paularena*, 2001]. We use ACE to determine SW plasma and IMF conditions, because Wind was located very far from the Earth-Sun line.

The duration of the quasi-radial IMF intervals was about 1.5 hours (Figure 1), 2 hours (Figure 2), and 14 hours (Figure 3), respectively. Here we should mention about geomagnetic activity as an internal factor affecting the magnetopause location [*Petrinec and Russell*, 1993; *Sibeck*, 1994]. As one can see in Figure 1-3, there were no geomagnetic storms during these three days (minimum value $Dst_{min}$ ~ -25 nT). Hence, the ring current effect, which would lead to an inflated magnetosphere, is negligibly small, if any. Therefore, we will rule out the *Dst* index from the following consideration. Auroral activity, represented by the *AE* index, was quiet on 16 July and 4 August with maximum value $AE_{max} \sim 150$ nT, while moderate auroral activity was observed on 8 August with $AE_{max} \sim 600$ nT. We will consider last event in relation with dayside magnetopause erosion due to the field-aligned currents.

Figure 4 shows THEMIS locations in the GSM coordinate system during time intervals from 1950 - 2037 UT on 16 July, 0400 - 0600 UT on 8 August 2007, and 0400 - 1200 UT on 4 August 2007. At the beginning of each interval, the five THEMIS probes were located in the subsolar region, moving outward in the string-of-pearls configuration with THB leading and THA trailing. Geotail was located in the duskside magnetosheath at GSM (6, 15, 1.5) $R_E$ on 16 July, in the nightside magnetosheath at GSM (-10, 24, -13) $R_E$ on 8 August, and inside the magnetotail at GSM (-23, 10, -12 $R_E$) on 4 August.

We compare clock angles of the magnetosheath and interplanetary magnetic fields observed by Geotail, ACE and THEMIS to estimate the time delay for SW propagation (Figure 5). We obtain a 43 min lag from ACE to Geotail on 16 July (Figure 5a). Taking into account the time for plasma to propagate from the THEMIS probes to Geotail, results in a 41.5 min time lag from ACE to THEMIS. On 8 August (Figure 5b), the SW propagation times from ACE to THEMIS and from ACE to Geotail was determined to be 38.5 min and 43.5 min, respectively. On the morning of 4 August, there was no spacecraft in the magnetosheath. We considered an interval from 1400 to 1900 UT when THA was located in the magnetosheath (Figure 5c). During this interval, THB magnetic field variations lagged those at ACE by 63 to 68 min, while a direct solar wind propagation technique yields a similar delay of ~65 min. For the interval from 0200 to 1200 UT on 4 August we suppose that the direct propagation technique is also reliable and, hence, the average time delay is estimated to ~63 min.

*Shue et al.* [1998] (hereafter Sh98) and *Chao et al.* [2002] (hereafter Ch02) provide reference models for the location of the MP and bow shock, respectively, as functions of solar wind conditions. Note that the bow shock predicted by the Ch02 model does not depend on the MP location. The Ch02 model predicts a decrease in the distance to the $Q_{\|}$ bow shock caused by a decrease in the fast magnetosonic velocity for small cone angles. Among a number of bow shock models, the Ch02 model demonstrates the highest prediction capabilities for a wide range of upstream conditions [*Dmitriev et al.*, 2003].

We also correct an aberration of up to 6° due to the Earth revolution around the sun and fluctuations in the SW direction. The correction was performed on point-by-point basis. The upstream and THEMIS data have been converted into aberrated GSM (aGSM) coordinates, in which the *X*-axis is aligned with the SW velocity [e.g. *Dmitriev et al.*, 2003]. In the aGSM coordinate system, the radial IMF is aligned with the SW flow and X-axis. SW dynamic pressure is calculated as $Pd = 1.67 \cdot 10^{-6} \cdot D \cdot V^2$ (in nPa), where *V* is bulk velocity (in km/s) and $D = N_p + 4N_\alpha$ (in cm$^{-3}$) is corrected SW density including a *He* contribution. The *He* content was nearly constant at 4~5% on 16 July and 8 August, and ~3% on 4 August. The total SW pressure *Psw* is calculated as a sum of the dynamic pressure, thermal proton pressure and magnetic pressures of the solar wind.

## 3. Geomagnetically Quiet Event on 16 July 2007

An interval of prolonged quasi-radial IMF at 1950-2037 UT on 16 July 2007 is presented in Figure 6. The SW and geomagnetic conditions are quiet: the SW velocity (~450 km/s) is stable, the SW pressure *Psw* varies slightly about 1.5 nPa, and IMF *Bz* is small (~-1 nT). Top panel in Figure 6 displays ion spectrograms from THEMIS ESA plasma instruments [*McFadden et al.*, 2008]. The presence of $Q_{\|}$ mode is supported by intense fluxes in the high-energy channels of ion spectrograms as well as by enhanced fluxes of energetic particles (not shown) observed by THB in the magnetosheath until ~2035 UT. The magnetosheath is identified as a region of relatively dense plasma with very wide energy spectrum of ions. Note that after ~2035 UT, the small cone angle is unreliable because of a different time shifting for the solar wind propagation in the trailing edge of interval. That shifting is associated with arriving of another solar wind structure led by a discontinuity, which propagation in the magnetosheath is observed by the THEMIS probes at ~2035 UT. Hence, we cut our consideration of the $Q_{\|}$ interval at 2035 UT.

At the beginning of the event at ~1950 UT, all the THEMIS probes except for THA are located in the magnetosheath. The innermost THA probe is inside the magnetosphere that is in good agreement with the Sh98 model prediction. From 1952 UT the MP starts to expand and reaches distances of >12.7 $R_E$, such that the outer probes THC, D, E enter inside the magnetosphere for a period of ~40 min. The expansion is large, THB observes the magnetopause at distances of ~2 $R_E$ above the Sh98 model prediction. Note that application of other magnetopause models gives similar result within one standard deviation σ (~0.5 $R_E$) for *Psw*≥1 and 2σ for *Psw*<1: all the models are unable to predict such distant magnetopause. We have to emphasize that the total SW pressure and IMF *Bz* are almost constant during that time and, thus, the expansion cannot be caused by variations of those parameters. It is reasonable to attribute the expansion to a decrease of the cone angle from ~30° to ~10° occurred at 1950 to 1953 UT.

The expanding magnetopause propagates outward from THE to THD with velocity of 26 km/s, then the MP decelerates to 9 km/s on its path from THD to THB. On average, the MP takes ~7 minutes to pass the distance of ~0.72 $R_E$ between THE and THB that gives the average speed of ~11 km/s. The MP velocities estimated by such method are presented in Table 1. The estimation error of about 15% is originated mainly from the limited ~3-sec time resolution of the magnetic field and plasma data and also due to uncertainties in determination of a moment when a probe crosses the MP current sheet.

The MP and adjacent magnetosheath plasma should move with similar velocities. Magnetosheath layer adjacent to the MP passes THEMIS probes during ~30 sec. In Figure 7 one can see that the ambient plasma in this layer moves outward mostly in X direction with the velocities of $V_x \sim 20$ km/s as measured by THE at 19h50m40s UT, ~2 to 10 km/s (THD, and C at 19h51m20s UT), and ~15 to 30 km/s as observed by THB at 19h58m00s UT. These values agree very well with the estimated MP velocities of 26 and 9 km/s (two upper rows in the Table 1). Thus our estimations are reasonable and we can conclude that within one error the MP velocities are consistent with the velocities $V_x$ of magnetosheath plasma adjacent to the MP. Magnetic field was measured by THEMIS/FGM instrument [*Auster et al.*, 2008]. During the MP crossings, the magnetospheric field, observed just inbound the magnetopause, is 2.4-time larger than the dipole value calculated from IGRF model. Such a value is expected from shielding effect of Chapman-Ferraro current. Note, the crossings observed at ~1951, ~1954, and ~1958 UT are caused by the outward MP moving, i.e. the magnetopause position is not of equilibrium. From the THB observations of the MP crossing at ~1958 UT (see Figure 6), one can see that the total pressure in the adjacent magnetosheath layer is slightly smaller than the *Ptot* in the magnetospheric boundary layer and there is a little jump from *Ptot* = 0.6 nPa in the magnetosheath to *Ptot* = 0.8 nPa in the magnetosphere**.** We suggest that this jump is owing to the MP moving outward to a new equilibrium position corresponding to lower pressure in the magnetosheath. From 2004 UT, when THB observes minimum magnetospheric field and *Ptot* ~ 0.6 nPa, the magnetopause starts to move back.

At 2011 - 2015 UT, the outermost probe THB observes a magnetosheath rebound, which is accompanied by an enhancement of cone angle from ~10° to ~35° and southward IMF from ~0 to –2 nT. According to the Sh98 model prediction, the small change of IMF *Bz* does not affect the magnetopause location. However, it is important to note that the geomagnetic field in vicinity of distant magnetopause is weak, ~20 to 40 nT. Because of that weak magnetic field, the expanded magnetopause is very sensitive to small variations of both major driving parameters (*Psw* and *Bz*) and other parameters affecting the bow shock and magnetosheath formation, such as the IMF cone angle. Probably, in the present case both effects of southward IMF and increasing cone angle are responsible for the inward magnetopause motion.

Magnetosheath rebound, observed by THB at 2020 - 2025 UT, is not accompanied by any substantial enhancement of the SW pressure or southward IMF. Even worse, the SW pressure decreases down to 1.3 nPa that should push the magnetopause outward that is not the case. In addition, we observe an enhancement of the cone angle up to >20°, which persists until the end of the interval at ~2034 UT. Hence, the observed dynamics of upstream parameters explains hardly the magnetosheath rebound at 2020 - 2025 UT as well as the magnetospheric rebound at 2025-2033 UT. There should be another process driving the magnetopause.

During the interval on 16 July 2007 we find variations of the magnetosheath and magnetospheric parameters over a wide range of timescales. We calculate thermal ion $Pi_{th}$ and electron $Pe_{th}$ pressures using 3-second data of reduced distribution from the THB/ESA instrument, which was operating in fast survey mode until 2027 UT, and then it was turned into slow mode. We also calculate the ion thermal pressures $Pi_{th}$ using data from full distribution, which has lower time resolution of ~1.5 minutes. One can see a good consistency between the two data products. The total magnetosheath pressure is obtained as a sum of $Pi_{th}$, $Pe_{th}$ and magnetic pressure *Pm*.

From 1950 to 2035 UT, the THEMIS probes observe 1~2-min oscillations of the total pressure in the magnetosheath as well as in the magnetosphere. Those specific quasi-regular variations clearly indicate oscillating MP motion. The multiple magnetopause crossings observed from 1951 to 2011 can be also considered as a result of a long-period (~10 min) MP undulation. Similar wavy motions (oscillations) of the MP were reported earlier as transient events [*Sibeck*, 1995; *Sibeck and Gosling*, 1996].

Based on the THEMIS observations, we can estimate the average MP location by two independent methods. In the first one, we assume nearly constant MP velocity of 9 km/s for propagation from THB to the new equilibrium location, i.e. during 6 min from 1958 to 2004 UT. Hence, we obtain that at 2004 UT the expanding MP approaches to a distance of ~12.85 $R_E$. The other method is based on the magnetopause model. As we can see in Figure 6, the magnetopause location is predicted much better when the Sh98 model is applied for the magnetosheath pressure *Ptot* and IMF *Bz*. The inconsistencies can be explained by the fact that the Sh98 as well as any other MP model has shortcomings at very low pressures. We consider the magnetosheath pressure *Ptot* = 0.6 nPa, detected by THB at 1958 UT, as a lower pressure limit and calculate the upper limit for the MP expansion of ~12.4±0.5 $R_E$. Thus, two different ways give similar estimations of the MP expansion.

After 2004 UT, the MP starts to move back and at 2012 UT approaches a distance somewhere between THD and THB, which are located at 12 $R_E$ and 12.5 $R_E$, respectively. Hence we can estimate the MP equilibrium location somewhere between 12.5 and 12.7 $R_E$ that is an average distance between the two extreme points of 12 ~ 12.5 $R_E$ and 12.85 $R_E$.

Considering upstream conditions, we do not find any substantial changes or quasi-periodic variations of the solar wind parameters, except for the cone angle. At the beginning (~1952 UT) the outward motion of MP is rather related to a fast decrease of the cone angle from ~30° to ~15°. This decrease is accompanied by a gradual decrease of the magnetosheath total pressure *Ptot* from 0.8 to 0.5 nPa, as observed by the THB probe. Here we point out that during the time interval of small cone angles (1952-2035 UT), the THB probe observes very low magnetosheath pressure, which is almost balanced by the magnetospheric pressure. This quasi-balance is clearly seen during the THB magnetopause crossings, which are revealed as significant jumps of all parameters except the total pressure across the MP. Inside the magnetosphere, the magnetic pressure (*Pm*) dominates and has a low value, consistent with MP distances of 12.3 – 12.9 $R_E$.

The total pressure in the magnetosheath *Ptot* is by a factor of 2 lower than the SW pressure *Psw* as indicated by a ratio *K* = *Ptot*/*Psw* in Figure 6. Near the magnetopause, the value of *Ptot* is found to be ~0.5 nPa. The total pressure in the low-pressure magnetosheath (hereafter LPM) is mainly contributed by the thermal pressure, a sum of ion $Pi_{th}$ and electron $Pe_{th}$ thermal pressures. The pressure of turbulent magnetic field *Pm* is very weak as observed by THB. Hence, the magnetosheath plasma *β* is high. We examined simultaneous Geotail observations of the post-noon magnetosheath (Figures 5a, and Figure 6) and also found weak magnetic field of ~5 nT, which is characterized by fast variations in the orientation and magnitude. Hence, in the dayside magnetosheath THEMIS and Geotail observed similar conditions proper for $Q_∥$ bow shock.

During the LPM mode, we do not find correlation for rapid (~minute) variations of the magnetosheath pressure $Ptot$ with the SW pressure $Psw$ and cone angle. We have to emphasize that the MP expansion associated with the LPM mode is observed by THEMIS for an unusually long time of ~45 min.

## 4. Disturbed Event on 8 August 2007

Figure 8 shows multiple magnetosphere encounters of THEMIS at unusually large distances of 13.5 to 14.5 $R_E$ accompanied by quasi-radial IMF at 0400 - 0600 UT on 8 August 2007. The SW conditions (Figure 2) were slightly disturbed: IMF $Bz$ varied between -2 and 1 nT, SW velocity was ~600 km/s and $Psw$ varied around 1.3 nPa. The THB/ESA instrument operated in the slow survey mode. The ion thermal pressures calculated for the full and reduced data products show good agreement in the magnetosheath/magnetosphere region.

As we see in Figure 2, the cone angle decreases below 30° after 0420 UT, and the quasi-radial IMF lasts for ~2 hours. In Figure 8 we can see that THA observes intense fluxes of energetic particles (>10 keV) and strong magnetosheath pressure variations indicated by the $K_A$ ratio. Those features confirm the presence of $Q_∥$ bow shock. It is interesting to note a decrease of the energetic particle fluxes and pressure variations at 0439 to 0444 UT when the cone angle increases up to 25° and conditions for the $Q_∥$ bow shock are broken.

From ~0330 to 0420UT, the IMF was mostly southward with $Bz$= -2 nT that caused substorm activity with $AE$ of ~600 nT, which continues until 0520 UT. Therefore, from 0420 to 0520 UT the magnetopause is driven by two opposite effects: the small cone angle and enhanced geomagnetic activity. Because of decreasing cone angle, one can expect an expansion of the magnetopause. Simultaneously, the substorm activity results in earthward motion of the dayside magnetopause because of a depression of the dayside geomagnetic field by the intensified field-aligned currents [*Sibeck*, 1994].

A response of the magnetopause and bow shock to the enhanced substorm activity is demonstrated in Figure 8. By ~0408 UT, all THEMIS probes were located inside the magnetosheath at distances of 13~14 $R_E$ that is in agreement with model predictions of the magnetopause and bow shock. After ~0408 UT, the outermost probes successively observes the bow shock moving inward and enter into the interplanetary medium, which is characterized by very narrow ion spectrum with mean energy of several keV. From 0418 to 0438 UT, the bow shock is located between THA and THE, somewhere at ~13.5 $R_E$, that is ~1 $R_E$ less than the Ch02 model prediction. The THEMIS encounter with interplanetary medium might result from the substorm-associated earthward motion of the dayside magnetopause, which is followed by the bow shock.

From ~0446 UT the SW pressure is gradually decreasing that leads to outward bow shock moving. The outermost THEMIS probes return to the magnetosheath at distance of ~14.2 $R_E$, which is close to the modeled bow shock location. At 0453 UT, the SW pressure has decreased down to ~1.2 nPa, the IMF $Bz$ starts to turn northward, and the substorm activity is weakening. At that time, the innermost THA observes a short (~1 min) magnetopause rebound at 13.5 $R_E$. It means that the magnetopause has expanded by more than 2 $R_E$ from the modeled location. During this crossing, an extreme LPM with $Ptot$ of 0.2~0.3 nPa (<30% of the SW pressure $Psw$~1.3 nPa) is observed by all the probes. At ~0457 to 0500 UT, the THEMIS probes successively cross the magnetopause, which is moving outward with velocity of ~25 km/s (Table 1) up to distances of ~14.5 $R_E$. Unfortunately, THEMIS did not provide high-resolution data on plasma velocities at that time.

The LPM pressure is balanced by the small pressure of magnetic field of ~20 nT in the magnetosphere. From 0500 to 0533 UT we can distinguish three magnetospheric intervals lasted for 4-8 min and recurred each 5-8 min. It is interesting to note that during the first and second intervals, when the $AE$ index is still high, the observed geomagnetic field is only 1.5-time higher (even not double) than the dipole magnetic field. The pressure balance during ±30 sec of those crossings almost conserves for the outward MP motions at 0500 and 0515 UT, when the MP passes THB. This balance indicates that the magnetopause would not move far away and stops near the THB orbit at ~14.5 $R_E$. On the other hand, for the observed minimal magnetosheath pressure of 0.16 nPa we can determine the modeled MP distance of ~15.7±0.5 $R_E$, i.e. ~1.3 $R_E$ above THB. These two features diminished geomagnetic field and smaller MP distance, can be attributed to a suppressing magnetic effect of the substorm activity at the restoring phase.

The magnetosheath encounter at 0506 - 0515 UT is accompanied with substantial increase of the cone angle. The MP moves very fast during this transient event (Table 1). At 0520 – 0525 UT, the THEMIS probes are located in the magnetosheath and observe enhanced plasma and magnetic pressure, and large negative $Bz$. It is rather difficult to determine unambiguously solar wind sources for those magnetosheath features. Hence, that magnetosheath rebound might be related to MP undulation with a period of ~10 min.

At 0525 – 0533 UT the SW pressure decreases to 1.1 nPa and the THEMIS probes reenter to the magnetosphere, where they observe magnetic pressure of 0.17 nPa. During the third magnetospheric interval, the $AE$ index decreases substantially and geomagnetic field approaches to the 2.4 dipole value. A minimum in the geomagnetic field profile at ~0529 UT indicates that the MP continues to move after the crossings and might reach even 16 $R_E$ against 11.5 $R_E$ predicted by the Sh98 model. Note that the model prediction is substantially improved by using the magnetosheath pressure $Ptot$ measured by the THA probe. We should note that the observed MP is located very close (within 0.5 $R_E$) to the bow shock predicted by the Ch02 model. It is very unlikely that the magnetosheath has so small thickness. Hence, we expect more distant bow shock during the LPM. We can estimate the magnetosheath thickness and bow shock distance from THEMIS observations of the magnetopause crossings at 0533 UT and bow shock crossings at 0535 UT. Using the time delay technique, we find that at 0533 UT, the MP moves inward with velocity of ~100 km/s (see Table 1). In a similar manner we can determine the velocity of bow shock of ~100 km/s at 0535 UT. Taking into account the 2-min time delay between the magnetopause and bow shock crossings, we estimate the path of ~1.9 $R_E$ passed by the bow shock until the crossing with THB. That path should be close to the thickness of magnetosheath. Hence, at 0533 UT the bow shock might be located at ~14.5 + 1.9 = 16.4 $R_E$ and the thickness of magnetosheath is estimated to be ~1.9 $R_E$. Such a thin magnetosheath is reported by *Jelínek et al.* [2010, submitted].

In the tail region, Geotail also observes an unusual MP expansion. The ion spectrograms presented in Figure 9 show that most of time Geotail is located in the magnetosheath, which is characterized by very variable magnetic field. During that interval, the LEP plasma instrument operated in a solar wind mode, which was switched to the magnetospheric mode only for a short time from 0520 UT to 0545 UT. At ~0523 to ~0530 UT, Geotail enters into the magnetosphere at very large distance of ~28 $R_E$ from the X-axis. At that time, the SW pressure is Psw~1.1 and the Sh98 model predicts the magnetopause distance of 21 $R_E$, i.e. ~7 $R_E$ smaller than the observed one. The magnetosphere encounter is revealed as a strong decrease of the ion density and enhancement of the magnetic field that are proper to conditions in the southern lobe/mantle. The surrounding regions, where the magnetic field magnitude is depressed and strongly fluctuating, can be attributed to the magnetosheath region downstream of the $Q_\parallel$ bow shock.

Here we have to point out very well correlation of the variations of magnetic field orientation (clock and cone angles) observed by Geotail and THA in the magnetosheath, and by ACE in the far upstream region (see Figures 5b, 8, and 9). The correlation is broken when magnetopause approaches to THA (at 0500 - 0530 UT) or to Geotail (at 0515 - 0530 UT). The coincidence of magnetosheath magnetic field orientation with the IMF orientation supports our suggestion that the magnetosphere is indeed affected by the solar wind structure with quasi-radial IMF as observed by ACE.

From ~0535 UT, the IMF is gradually turning southward, the SW pressure is increasing up to 1.2 nPa and the cone angle is varying about 25°~30°. The THEMIS probes approach to apogee of 14.7 $R_E$ and return to the magnetosheath and/or bow shock region.

Thus, during this prolonged expansion event (about 40 min~1 hour) we reveal significant differences between the observed MP location and the Sh98 model: ~3.5 $R_E$ in the dayside and ~7 $R_E$ in the tail region. The observed magnetosheath pressure near the magnetopause was ~0.16 nPa and the ratio $K$~0.2, both are extremely small. The dayside MP undulates with a period of ~10 min near a new equilibrium position, which we find at ~13.5 to 14.5 $R_E$, i.e. somewhere between the innermost THA and outermost THB probes. In the beginning of interval considered, that equilibrium MP location is substantially affected by the enhanced substorm activity.

**5. Long-Lasting Event on 4 August 2007**

A 14-hour interval of quasi-radial IMF occurred at 0100 to 1500 UT on 4 August 2007. As one can see in Figure 3, the event is characterized by steady and quiet SW and geomagnetic conditions: the SW velocity (~400 km/s), the SW total pressure is low and decreases from 0.7 nPa to 0.5 nPa, IMF $Bz$, $AE$ and $Dst$ are small. The models predict the MP and bow-shock location at ~12.5 $R_E$ and 17~18 $R_E$, respectively (see Figure 10).

In Figures 3 and 10 we find the expanded MP observed by the outer THEMIS probes continuously during ~4 hours from ~0300 to ~0700 UT. Then until ~0800 UT they observe magnetosheath intervals with of a few minutes of duration. After that time, when THEMIS approaches to apogee of ~14.7 $R_E$, the probes enter deep into the magnetosheath and sometimes encounter with the magnetosphere.

Figure 10 demonstrates a part of that at 0400 - 1200 UT, when the THEMIS probes are located at the distances from ~12 to 14.7 $R_E$ (see Figure 4c). During 0400 - 0700 UT, all the probes observe the magnetosphere. However, the magnetopause model predicts magnetosheath for the outer probes THB, C, D, and E that is not the case. Since THB magnetic field on average is 2.5 times stronger than the dipole, we infer that THB located at ~14.4 $R_E$ observes shielding effect of the Chapmen-Ferraro current, and, hence, it is close to the magnetopause. That is supported by multiple MP crossings observed by THB at 07-08 UT.

In the magnetosphere, THEMIS probes observe quasi-periodic variations of the geomagnetic field with average period of ~10 min that indicates MP undulations. Sometimes, about one time per hour, the fluctuations of MP location are so large that the THB crosses the magnetopause. Transient magnetosheath rebounds of ~1 min duration are observed by THB at ~0430 UT, ~0525 UT and ~0630 UT. Note, that during this 3-hours interval, we find no obvious correlation of the magnetospheric field variations with the SW pressure, though a prominent change of the SW pressure at ~0500-0520 UT does produce geomagnetic field decrease.

After 0700 UT, the outer probes approach to the magnetopause and observe multiple magnetosheath encounters. The innermost THA probe does not leave the magnetosphere until ~0810 UT and observes geomagnetic field variations correlating well with the inward and outward magnetopause motion. We have to point out that those MP fluctuations as well as others occurred later (see, for example, THA at 09-10 UT) do not relate to variations of solar wind parameters. Similar situation is revealed for three magnetosphere rebounds observed by all the probes at 0820 - 0840, 0930 - 0945 UT and at 1040 - 1110 UT. Moreover, the MP crossings of THA do not correlate with the magnetosheath pressure variations observed by the outer THEMIS probes.

From ~0810 UT all the THEMIS probes successively enter the magnetosheath. The innermost probe THA crosses the magnetopause at distance of ~14.0 $R_E$ and enters the magnetosheath for 5 min. The average velocity of the inward MP motion is estimated of ~16 km/s that is not typical for transient events. From 0820 all satellites are located inside the magnetosphere and observe decreasing geomagnetic field with minimum at ~0830 UT. It means that the MP moves far from the outermost THB probe located at 14.6 $R_E$, i.e. the magnetopause is at distances, which are at least ~2.1 $R_E$ larger than the Sh98 model prediction of ~12.5 $R_E$. The model prediction becomes much more accurate when we use the magnetosheath total pressure *Ptot* measured by THB instead solar wind pressure *Psw*.

From 0840 UT, all probes enter to the magnetosheath. Comparing THE and THA locations and magnetospheric field profiles from 0830 to 0900 UT we determine the MP velocity of 5 km/s and average MP position between 14.2 and 14.5 $R_E$. The magnetosheath intervals at 0840 - 0930, 0945 - 1040 and after 1110 UT are highly turbulent and populated by sporadic structures of high plasma pressure, which are similar to magnetosheath transient plasma jets [*Němeček et al.*, 1998; *Savin et al.*, 2004; 2008]. Such transient jets are characterized by intense localized ion fluxes, which kinetic energy density can be even higher than those in the upstream solar wind.

In the present case, the magnetosheath total pressure measured by THB fluctuates from 0.3 to 0.7 nPa, and the ratio $K$ varies quickly between 0.3 and 1.3. The LPM is characterized by quasi-static flow balance with the base line of $Psw$~0.3 nPa and $K$ ~ 0.4 to 0.5. That balance is disturbed by inherent transient dynamics manifested in the plasma jets. There is no obvious correlation of the magnetosheath pressure variations with the dynamics of cone angle and/or SW pressure.

## 6. Discussion

We have analyzed three cases of quasi-radial IMF and revealed substantial magnetopause expansions accompanied by nearly constant solar wind total pressure. With in situ THEMIS and Geotail observations, we have found that during quasi-radial IMF, the whole magnetosphere is expanded significantly, far beyond the expected position. Dramatic decreases in the magnetosheath total pressure in the each case were observed by the THEMIS probes.

At ~0525 on 8 August, THEMIS observed the subsolar magnetopause at distance of ~14.5 $R_E$ that is >3 $R_E$ distant from the model prediction. At the same time, Geotail observed the MP in the tail region at distances of ~7 $R_E$ far from the model prediction of ~21 $R_E$ for $Psw$ ~ 1.1 nPa. That is different from the assumption of "bullet-like" magnetopause proposed by *Merka et al.* [2003]. The maximum magnetopause distance of 14.7 $R_E$, observed by THEMIS in apogee at ~1100 UT on 4 August for $Psw$ ~ 0.6 nPa, is restricted by the orbital bias. We estimate that the subsolar magnetopause might expand up to 16 $R_E$.

Such a distant position is proper to the bow shock rather than to the magnetopause. Because of the orbital bias, the distant bow shock could not be observed for those cases. Based on average velocities of the magnetopause and bow shock observed at 0533 - 0535 UT on 8 August, we estimate the bow shock distance of ~16.4 $R_E$ and magnetosheath thickness of ~1.9 $R_E$ that is substantially different from their nominal values of ~15 $R_E$ and ~4 $R_E$, respectively. This discrepancy is a subject of further investigations based on THEMIS data in 2008 to 2009 when the outer probes move to larger distances from the Earth.

For quasi-radial IMF, we have found an ambiguous dependence of the subsolar magnetopause distance on the solar wind pressure. Namely, the average location of the expanded subsolar magnetopause is estimated of ~12.5 to 12.7 $R_E$ for the SW pressure $Psw$ ~ 1.3-1.5 nPa at ~2000-2030 UT on 16 July; ~14.5 $R_E$ for the $Psw$ ~ 1.1-1.3 nPa at ~0500-0530 UT on 8 August; and ~14.4 $R_E$ for the $Psw$ ~ 0.5-0.6 nPa at ~06-08 UT on 4 August. The difference in the magnetopause locations can not be explained by the effect of southward IMF because the magnitude of IMF $Bz$ was very small. These cases did not accompanied by geomagnetic storms. Hence, the magnetopause location in these cases is controlled by other driving parameters.

A significance of these driving parameters is demonstrated in the following example. Comparing Figures 6 and 8, we reveal that for the same SW pressure of ~1.3 nPa and northward IMF, the subsolar MP is located between THA and THE (i.e., between 10.5 $R_E$ and 11.7 $R_E$) at ~1955 UT on 16 July, whereas it is beyond 14.5 $R_E$ at ~ 0530 UT on 8 August. From the above we can determine the maximal observed displacement of the subsolar MP of at least >3 $R_E$, and possibly as large as ~5 $R_E$ that corresponds to ~30% uncertainty in the MP location.

It is well known that the magnetopause is driven directly by the plasma and magnetic pressure of adjacent magnetosheath. According to classical hydrodynamic theory [see *Spreiter et al.*, 1966 for reference], a ratio $K$ of the stagnation pressure at the subsolar magnetopause to the upstream SW pressure should approach to 0.881 when Mach number is much greater than unity. However, after the late 80's scientists found indications that MP location under quiet condition (northward IMF) is controlled not only by the SW pressure but also by the IMF orientation [e.g., *Fairfield et al.*, 1990; *Sibeck*, 1995]. It was proposed that during radial (transverse) IMF, the pressure applied to the magnetopause is smaller (higher). This idea was used for interpretation of unusually distant MP [*Laakso et al.*, 1998; *Merka et al.*, 2003; *Zhang et al.*, 2009].

THEMIS observations of the low-pressure magnetosheath support the idea proposed by *Fairfield et al.* [1990]. The fraction of the SW pressure applied to the magnetopause depends upon the orientation of IMF, and for radial IMF the ratio $K$ is considerably smaller than theoretical prediction of 0.881. In the case of pronounced LPM, we discover very low thermal pressure $P_{th}$ and extremely low magnetic pressure $P_m$ in the magnetosheath, such that only a small portion of solar wind kinetic energy is applied to the subsolar magnetopause and the ratio $K$ is ~0.5 and even less. Under such conditions, the magnetosheath plasma $\beta$ is very large. The high-$\beta$ magnetosheath for quasi-radial IMF was reported by *Le and Russell* [1994]. They showed that during quasi-radial IMF the high value of plasma $\beta$ in the magnetosheath does not depend on the IMF strength and value of the solar wind plasma $\beta$.

We have to note that the accuracy of ratio $K$ calculation can be strongly affected by the quality of upstream solar wind data and THEMIS calibration errors. It is known that the characteristics of SW plasma and IMF affecting the magnetosphere might be different from those observed far upstream of the Earth [*Zastenker et al.*, 1998; *Richardson and Paularena*, 2001; *Riazantseva et al.*, 2002]. The difference increases with a spacecraft separation perpendicular to the sun-earth line (P-separation), as we can see in Figures 1, 2, 3. In order to minimize this effect we use upstream data provided by the ACE monitor, which has the smallest P-separation. In addition, the solar wind with small IMF cone angles is more structured than that for the perpendicular IMF, such that even for small P-separation, solar wind structures observed far upstream correlate weakly with those observed near the Earth. Without a near-earth satellite, this effect is difficult to rule out.

However, major parameter controlling that correlation is a variability of the SW density. In Figures 1, 2, 3, one can see relatively weak density variations as observed by ACE. Under such conditions the SW dynamic pressure detected by ACE is close to that detected by Wind at very large P-separation. Hence, it is unlikely that the solar wind plasma conditions, affecting the Earth's magnetosphere, appear substantially different than that observed in a wide spatial range by ACE and Wind. At the same time, we can point out a pure correlation for the IMF such that the quasi-radial IMF is observed by Wind occasionally. It is reasonable to suggest that Wind observes different IMF due to the large P-separation. Hence, the ACE plasma and magnetic data are more reliable for the present study.

Absolute calibration of the THEMIS plasma instruments was done through cross-calibration with the Wind-SWE instrument [*McFadden et al.*, 2008]. As for considered period of summer 2007, the authors also executed a final test of the absolute calibration, when magnetopause crossings were evaluated to check for pressure balance, i.e. the same way as we use in our study. The total pressure was found to be nearly constant during the MP crossings that proves the accurate absolute calibrations of the plasma instruments. Here we have to point out that the THEMIS/ESA instrument operates in various modes. The high-resolution measurements of plasma parameters, including velocity, are provided in the fast and slow survey full modes with 1.5 min and 6 min resolution. Very often in the magnetosheath the instrument operates in the fast survey reduced mode with a low angular/energy resolution and high time resolution (3-sec). As a result, that mode provides reliable data only for low-speed plasma. In the first case event on Jul. 16, the THEMIS/ESA operated in the fast reduced mode. In Figure 7 we show that the transversal components of the magnetosheath plasma velocity are not very large in close vicinity of the magnetopause. Hence, the most reliable plasma data and total magnetosheath pressure can be obtained only near the magnetopause. The second and third case events, when plasma velocity was unavailable (slow reduced mode with omni-directional spectra), were analyzed on the base of that rule. By this way we obtain reliable estimations of the ratio $K$ derived from the ACE and THEMIS data.

The problem of solar wind energy transformation in the LPM mode is an important but still poorly understood issue. It is quite possible that the origin of LPM is related to particular formation of the bow shock and magnetosheath under quasi-radial IMF conditions that result in redistribution of the solar wind energy and decreasing the portion of energy affecting the magnetopause. First of all, the transverse component of quasi-radial IMF is so small that magnetic field is weakly amplified at the bow shock and in the magnetosheath [*Le and Russell*, 1994].

In the literature we have found a few mechanisms, which might cause a low ratio $K$. *Wilkinson* [2003] represents the high-Mach-number $Q_\parallel$ bow shock as a thick (≥2-2.5 $R_E$, radially) magnetic pulsation region, characterized by ion reflection at bow shock front and leakage from the magnetosheath with propagation far into upstream region and by a rich variety of interacting wave modes and particle distributions. In that region the SW is heated and deflected often by 20-40° or more. *Schwartz and Burgess* [1991] propose a general description of that transition zone as a quite filamentary three-dimensional structure. Deceleration of the solar wind upstream of the $Q_\parallel$ bow shock is also essential due to the interaction with short large-amplitude magnetic structures (SLAM) [*Schwartz et. al*, 1992] and with ion foreshocks [*Zhang et al.*, 1995].

*Savin et al.* [2008] suggested another mechanism of the solar energy redistribution inside the magnetosheath. They found that the magnetosheath kinetic energy density during more than one hour can exhibit an average level and a series of jets, i.e. peaks far exceeding the kinetic energy density in the undisturbed solar wind. It was suggested that dynamic interaction in the magnetosheath plasma is non-uniform and intrinsically transient, as the plasma is still evolving from the shocked to a statistically equilibrium turbulent state. In the course of this evolution, it seems that processes may occur which concentrate the free energy in the still underdeveloped turbulence and focus the plasma into jets. It was noted that the jets could weakly interact with the magnetopause and, thus, provide the super-diffusive plasma transport inside the magnetosphere. Apparently, in the presence of jets, the background magnetosheath energy should be decreased.

SW structures with quasi-radial IMF are observed quite often at declining speed profiles within the trailing portions of ICME [*Neugebauer et al.*, 1997] or within corotating rarefaction regions [*Jones et al.*, 1998; *Gosling and Skoug*, 2002]. Those structures, expanding from the Sun, can last from hours to several days. They are characterized by relatively weak IMF and relatively low plasma density/pressure in the upstream solar wind [e.g. *Riley and Gosling*, 2007].

In order to estimate numerically the characteristic properties of solar wind for quasi-radial IMF in the 23$^{rd}$ solar cycle, we have performed a statistical analysis of 16-sec ACE magnetic and 1-min plasma data for 11 years from 1998 to 2008. In Figure 11 a statistical distribution of the solar wind dynamic pressure, measured by ACE during intervals of quasi-radial IMF, is compared with common distribution for 11 years. A deficiency of medium and high pressures is revealed for the intervals of quasi-radial IMF. The mean pressure for those intervals is ~1.4 nPa that is smaller than the mean of 1.7 nPa for the common distribution. Note that the mean pressure of 1.7 nPa is smaller than the average SW dynamic pressure of 2 nPa obtained for four solar cycles. That relatively small mean pressure results from relatively low solar wind density of ~2 to 4 cm$^{-3}$ owing to an abnormal behavior of the 23$^{rd}$ solar cycle [*Dmitriev et al.*, 2009]. Therefore, the MP expansion related to quasi-radial IMF can be masked by the effect of low solar wind pressures that makes difficult statistical finding of the quasi-radial IMF effect.

From the statistical analysis we also find that cone angles of <30° are observed for ~16% of time. Figure 12 shows statistical distributions of integral occurrence probability of duration of intervals with cone angles below 30° for whole 11-year period and for one year in solar minimum. One can see that the intervals with duration more than 10 min contribute to ~30% of statistics. Hence, they can be observed for ~5% of time. 5-min intervals occur for ~8% of time. The number of long-lasting intervals is higher in the solar minimum. Thus, the quasi-radial IMF occurs quite often. In this sense, the phenomenon of LPM-associated MP expansion might be rather typical than unusual and, thus, the effect of small cone angle should be taken into account in future magnetopause modeling.

Figure 13 illustrates the effect of magnetopause expansion for the LPM mode. The MP crossings observed by THEMIS on 16 July and by THEMIS and Geotail on 8 August 2007 can be predicted by the reference model applied for the magnetosheath pressure of 0.5 nPa ($K$=0.3) and 0.1 nPa ($K$=0.07), respectively. Note, that the SW pressure for those cases was ~1.1-1.5 nPa. The MP crossings observed by THEMIS on 4 August 2007 are well described by the model calculated for the magnetosheath pressure of 0.3 nPa, while the SW pressure is 0.6 nPa.

The expanded outer magnetosphere has a lower magnetic field and, thus, becomes more sensitive to variations of both major and minor driving parameters. As a result, a small change in the SW pressure and/or IMF orientation can lead to a substantial transient motion of the boundary. We have also found MP displacements in response to variations of substorm activity, represented by the $AE$ index. Therefore, during LPM the effect of cone angle can strongly interfere with effects produced by other driving parameters.

We observed several cases of prominent MP inward/outward motion when the cone angle exceeds/falls down a certain threshold of 20 to 25°. However, we also have found a number of cases when the MP motion is not related to both variations of the upstream parameters, including the cone angle, and the magnetosheath pressure. Such motion probably can be attributed to the MP undulations with a wide range of periods. Thus, a feature of the MP dynamics for long-lasting quasi-radial IMF is characterized by a superposition of the steady-state expansion and wavy MP motion. New equilibrium position of the MP can be remote by several $R_E$ from the nominal. That position is mainly controlled by the ratio $K$, which is much smaller than the theoretical prediction of 0.881. The magnetopause undulates near new equilibrium location. The velocity of undulating magnetopause is found to be highly variable from several km/s to >200 km/s (see Table 1). Similar range of the MP velocities for quasi-radial IMF was reported by *Le and Russell* [1994].

We have to point out that the ratio $K$ has no direct linear relationship with the cone angle. We observe that for the large cone angles of >25°, the ratio increases and approaches to its theoretical value. However, the small cone angles (<20°) are accompanied by the $K$ varying in a wide range from 0.16 to 0.6. We can assume that the value of $K$ for quasi-radial IMF depends on the upstream SW plasma $\beta$. On 16 July and 4 August, when the proton $\beta$ was much smaller than 1, the value of $K$ was about 0.5. During the interval of very low $K$ on 8 August, the SW plasma $\beta$ was close to 1 and even larger.

Our assumption is based on results of magnetosheath modeling. *De Sterck and Poedts* [1999, 2001] investigated the bow shock and magnetosheath topology for quasi-radial IMF, Mach number less than 6 and low proton $\beta$ (<0.6). The 3-dimensional MHD simulation was performed for the idealized setting of flow around a rigid paraboloid magnetopause. The authors reveal very complex topology of the bow shock and magnetosheath, which is controlled by three SW parameters: $\beta$, Mach number and IMF cone angle. It is hard to apply those results directly to our cases, which are accompanied by high Mach number and relatively high proton $\beta$ (>0.6). However, it is possible that the same driving parameters might control the LPM mode.

In the present study we demonstrate three cases, which are characterized by different durations, upstream solar wind and magnetosheath plasma properties and magnetospheric condition. But they have one common feature of LPM. It's quite possible that the LPM might result from different mechanisms. Thus, we believe that further comprehensive statistical study of the magnetosheath plasma and magnetic field properties is important key to a clear insight into the mechanisms of the LPM formation.

# 7. Conclusions

With THEMIS data, we reveal that the magnetopause expansions are caused by significant decrease of total pressure in high-$\beta$ magnetosheath (LPM mode). Prominent LPM mode is observed when the IMF cone angles are less than 20 ~ 25°.

From simultaneous observations of Geotail and THEMIS, we infer a global expansion of the magnetopause. The magnetopause is found more than 3 $R_E$ and ~7 $R_E$ far from the nominal location in the dayside and tail region, respectively.

The MP expansion can persist for a few hours, as long as quasi-radial IMF conditions, that indicates a steady-state process driving the magnetopause.

The equilibrium MP position was determined at 12.5 to 12.7 $R_E$ for the upstream SW pressure $Psw$ ~ 1.3-1.5 nPa and the adjacent magnetosheath total pressure $Ptot$ ~0.5 nPa; ~14.5 $R_E$ for $Psw$ ~ 1.1-1.3 nPa and $Ptot$ ~0.16-0.3 nPa; and ~14.4 $R_E$ for $Psw$ ~ 0.5-0.6 nPa and $Ptot$ ~0.25-0.35 nPa. The equilibrium MP position is affected by geomagnetic activity.

Minimal value of the total pressure observed by THEMIS in the adjacent magnetosheath is 0.16 nPa and, thus, the fraction $K$ of the SW pressure applied to the MP can be as extremely small as 0.2. The ratio $K$ decreases with increasing upstream SW plasma $\beta$.

Statistical study of 11 years of ACE data reveals that the quasi-radial IMF conditions are not very rare and occur for ~16% of time. Those conditions frequently interfere with the small solar wind pressure that makes difficult to distinguish the cone angle effect statistically.

**Acknowledgements** We acknowledge NASA contract NAS5-02099 and V. Angelopoulos for use of data from the THEMIS mission. We thank K. H. Glassmeier, and U. Auster for the use of FGM data provided under contract 50 OC 0302. We thank N. Ness and D. J. McComas for the use ACE solar wind data made available via the CDAWeb. The Geotail magnetic field and plasma data were provided by T. Nagai and Y. Saito, respectively. This work was supported by grants NSC 98-2811-M-008-043, NSC-98-2111-M-008-019 and NSC 99-2111-M-008-004. The work at Charles University was supported by the Czech Grant Agency under Contract 205/09/0112 and by the Research Plan MSM 0021620860.

**Table 1.** Magnetopause and plasma velocities

| 16 Jul | UT | probes | $V_{MP}$ (km/s) |
|---|---|---|---|
| 1 | 19:52 | E-D | 26 ± 7 |
| 2 | 19:58 | D-B | 9 ± 1 |
| 3 | 20:32 | B-D | -55 ± 4 |
| 4 | 20:33 | D-E | -14 ± 3 |
| 5 | 20:37 | E-A | -35 ± 4 |
| 08 Aug | 05:00 | A-B | 25 ± 5 |
| 2 | 05:05 | B-E | -105 ± 30 |
| 3 | 05:07 | E-A | -33 ± 5 |
| 4 | 05:11 | A-E | 180 ± 50 |
| 5 | 05:14 | E-B | 8 ± 1 |
| 6 | 0520 | B-A | -48 ± 3 |
| 7 | 0525 | A-B | 230 ± 80 |
| 8 | 0532 | B-A | -100 ± 10 |

**Figure captions**

**Figure 1.** Upstream solar wind parameters observed by ACE at 1600-2400 UT on 16 July, 2007 (from top to bottom): velocity components $Vx$ (gray line), $Vy$ (thick black line) and $Vz$ (thin line); proton density $D$ (thin) and temperature $T$ (gray); SW dynamic pressure $Pd$ observed by ACE (black) and Wind (gray); IMF strength $B$ (black) and $Bx$-component (gray); IMF component $By$ (gray) and $Bz$ (black); IMF clock (Cl) and cone (Ca) angles observed by ACE (black) and Wind (gray). Two bottom panels show geomagnetic indices $AE$ (black) and $Dst$ (SYM-index) (gray); and distances to the THB probe (solid line) and THA probe (dotted line), *Shue et al.* [1998] magnetopause model prediction (gray line), magnetosphere intervals observed by THB (black bars) and by THA (shaded bars). The time of upstream parameters is delayed on SW propagation to THEMIS (see explanation in the text).

**Figure 2.** The same as in Figure 1 but on 8 August, 2007 at 0200-0800 UT.

**Figure 3.** The same as in Figure 1 but on 4 August, 2007 at 0000-2400 UT.

**Figure 4.** GSM coordinates of the THEMIS probes for the time intervals: (a) 1950-2037 UT on July 16 2007; (b) 0400-0600 UT on August 8 2007; (c) 0400-1200 UT on August 4 2007.

**Figure 5**. Clock angle of magnetic fields: (a) on 16 July observed by Geotail (gray line) and ACE (black line) delayed by 43 min; (b) on 8 August observed by THA (gray line), Geotail (circles) delayed by -5 min and ACE (black line) delayed by 38.5 min; (c) on 4 August observed by THA (gray line) and ACE (black line) delayed by 63 min.

**Figure 6**. Plasma and magnetic fields observed on 16 July 2007 (from top to bottom): THEMIS ion spectrograms; *Chao et al.* [2002] bow shock model prediction (Ch02); *Shue et al.* [1998] magnetopause model predictions (Sh98) calculated for the solar wind $Psw$ (circles) and magnetosheath $Ptot$ (diagonal crosses) pressures; THEMIS radial distances (thick segments mark the magnetosphere encounters); ACE and THB measurements of magnetic field strength and $Bz$ (divided by 10 for THB); THB plasma velocity (Vtot) and components $Vx$, $Vy$, $Vz$; the upstream solar wind pressure $Psw$ and THB magnetic ($Pm$), thermal ion $Pi_{th}$, thermal electron $Pe_{th}$, and total pressure ($Ptot$), circles depict the ion pressure $Pi_{th}$ in ESA full mode; solar wind proton $\beta$ and ratio $K$ ($Ptot/Psw$); cone angles of ACE and Geotail magnetic field delayed by 41.5 and -1.5 min, respectively. Time intervals of THB magnetosphere encounters are marked by blue shadow bars.

**Figure 7**. Components $Vx$, $Vy$, $Vz$ of plasma velocity observed by THEMIS probes (THB,C,D,E) on 16 July, 2007 near the magnetopause crossings (indicated by vertical dashed lines) during transition from the magnetosheath to the magnetosphere at 1950-2000 UT.

**Figure 8.** The same as Figure 6, but for 0400 to 0600 UT on 8 August 2007. Instead of the panel with THB velocity components, we show a panel with $AE$. Ratio $K$ is shown for THA ($K_A$) by black line and for THB ($K_B$) by red line. The ACE and Geotail magnetic field cone angles are delayed by 38.5 and -5 min, respectively.

**Figure 9.** Geotail observations in the tail region on 8 August 2007 (from top to bottom): CPI and LEP plasma ion spectrograms, and magnetic field strength (Geotail time). The blue shadow bar indicates the magnetosphere encounter.

**Figure 10.** The same as Figure 6, but for 0400 to 1200 UT on 4 August 2007. A panel with THA magnetic field (magnitude and components) is shown instead of a panel with THB velocity components.

**Figure 11.** Statistical distributions of the solar wind dynamic pressure $Pd$ observed by ACE for quasi-radial IMF with cone angle <30° (black solid histogram) and for whole time (gray dotted histogram) in 1998 to 2008. The mean, median and most probable values of $Pd$ for those two distributions are about 1.4 and 1.7 nPa, respectively. A deficiency of medium and high pressures is revealed for the statistics of quasi-radial IMF.

**Figure 12.** Integral occurrence probability of intervals with quasi-radial IMF (cone angle <30°) constructed on the base of 16-second resolution ACE magnetic data for 11 years from 1998 to 2008 (black solid line) and for the year 2007 (gray dotted line). The 11-year distribution can be fit by a power function (dashed line) with the exponent of ~1.1. The solar minimum in 2007 is enriched by long-lasting intervals of quasi-parallel IMF.

**Figure 13.** GSM locations of the THEMIS probes and Geotail at 1950 - 2037 UT on 16 July 2007, at 0400 - 0600 UT on 8 August 2007, and at 0400-1200 UT on 4 August 2007. The MP profiles are predicted by a reference model [*Shue et.al.*, 1998] for various pressures: 0.1 (dashed line), 0.5 (dotted line) and 1.5 nPa (solid line).

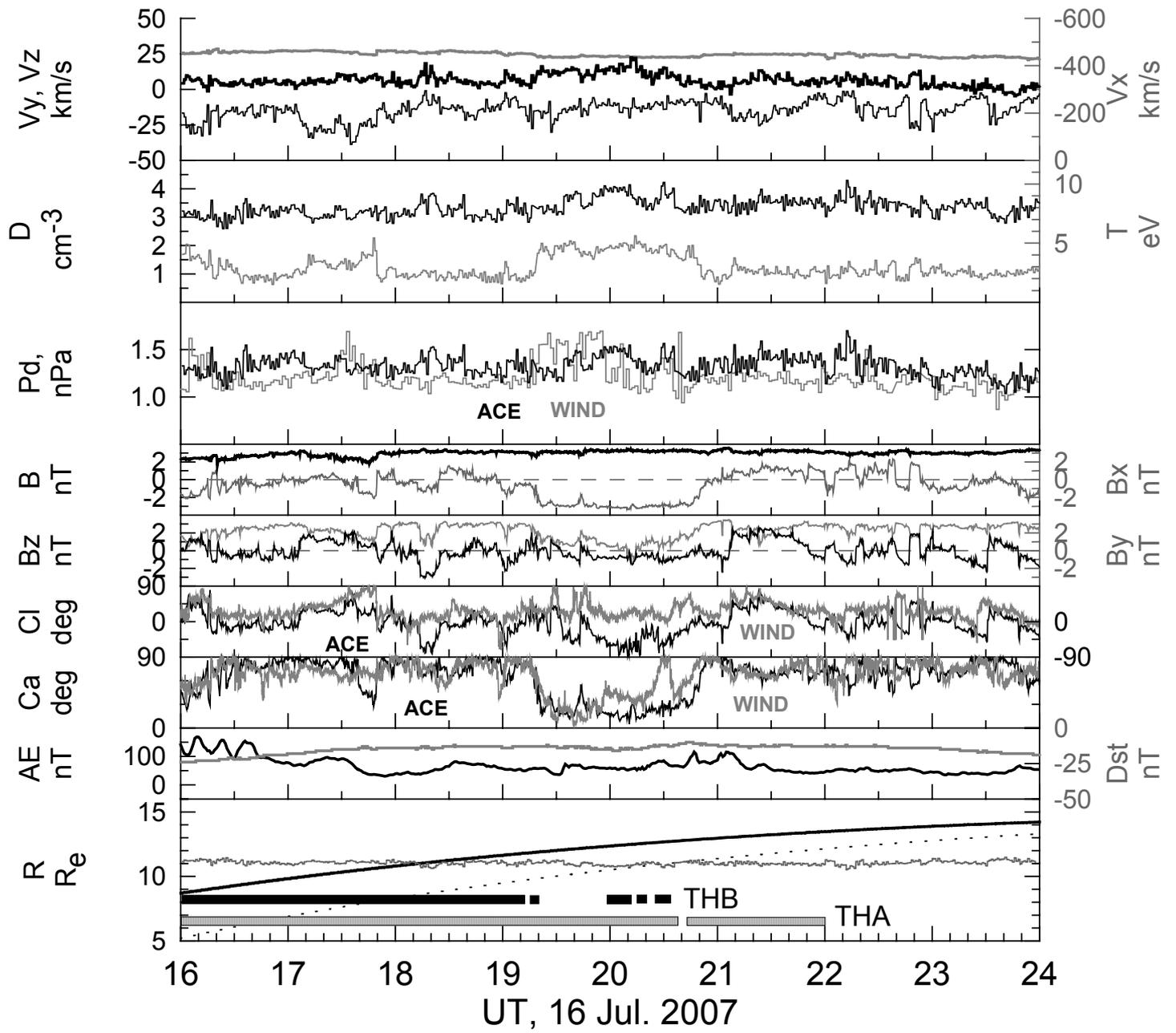

Fig.1

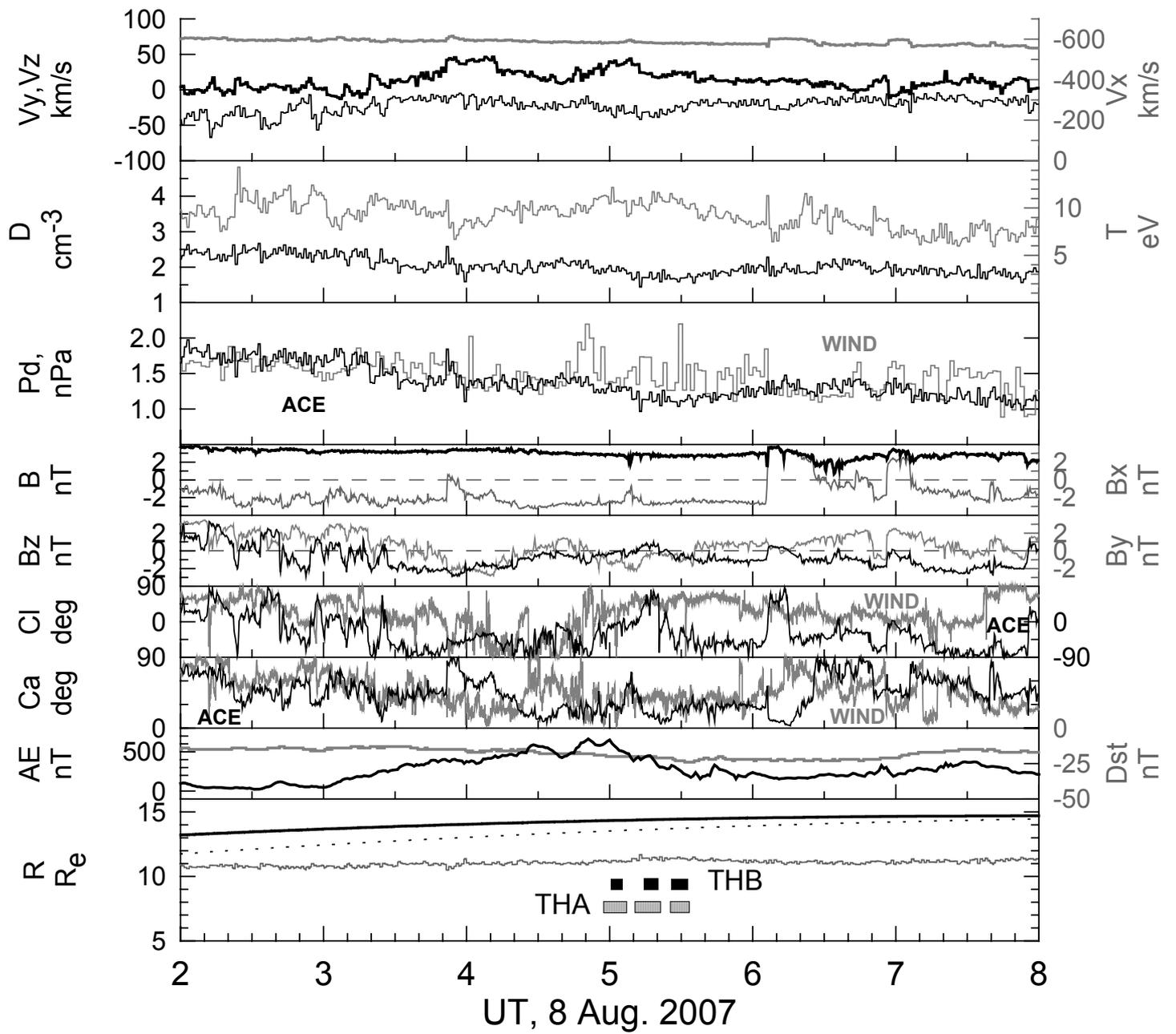

Fig.2

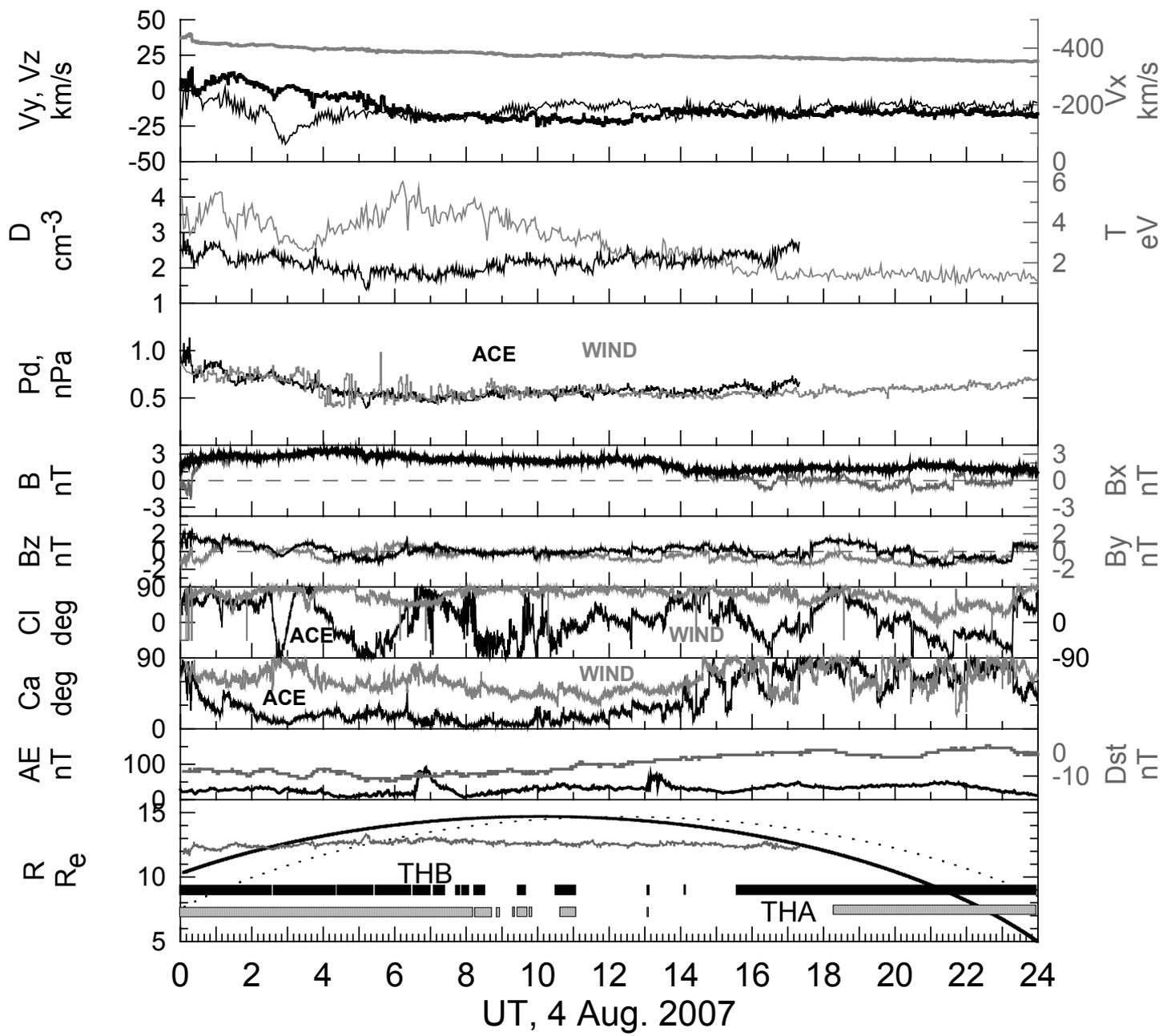

Fig.3

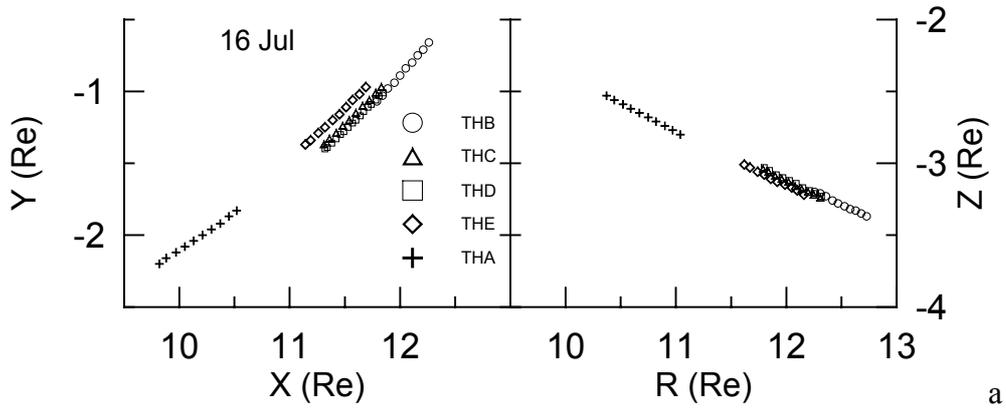
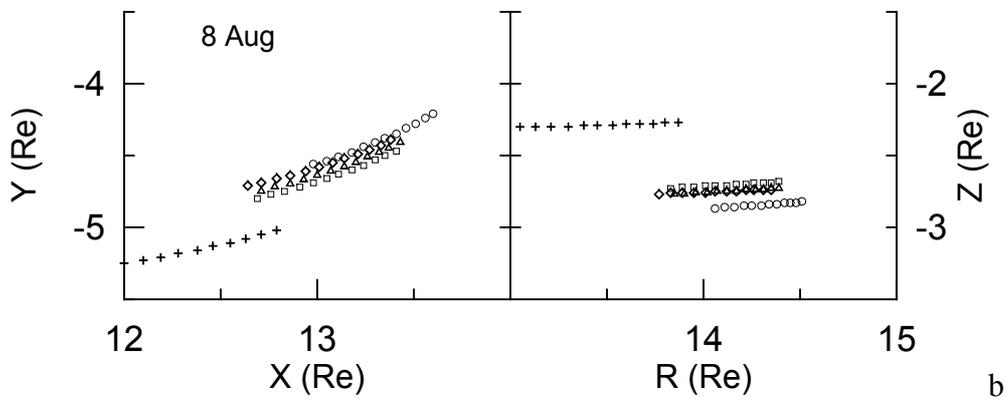
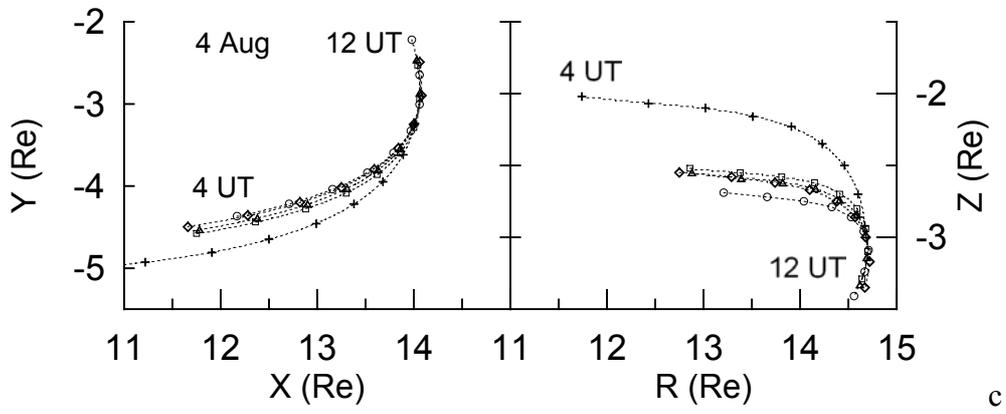

Fig.4

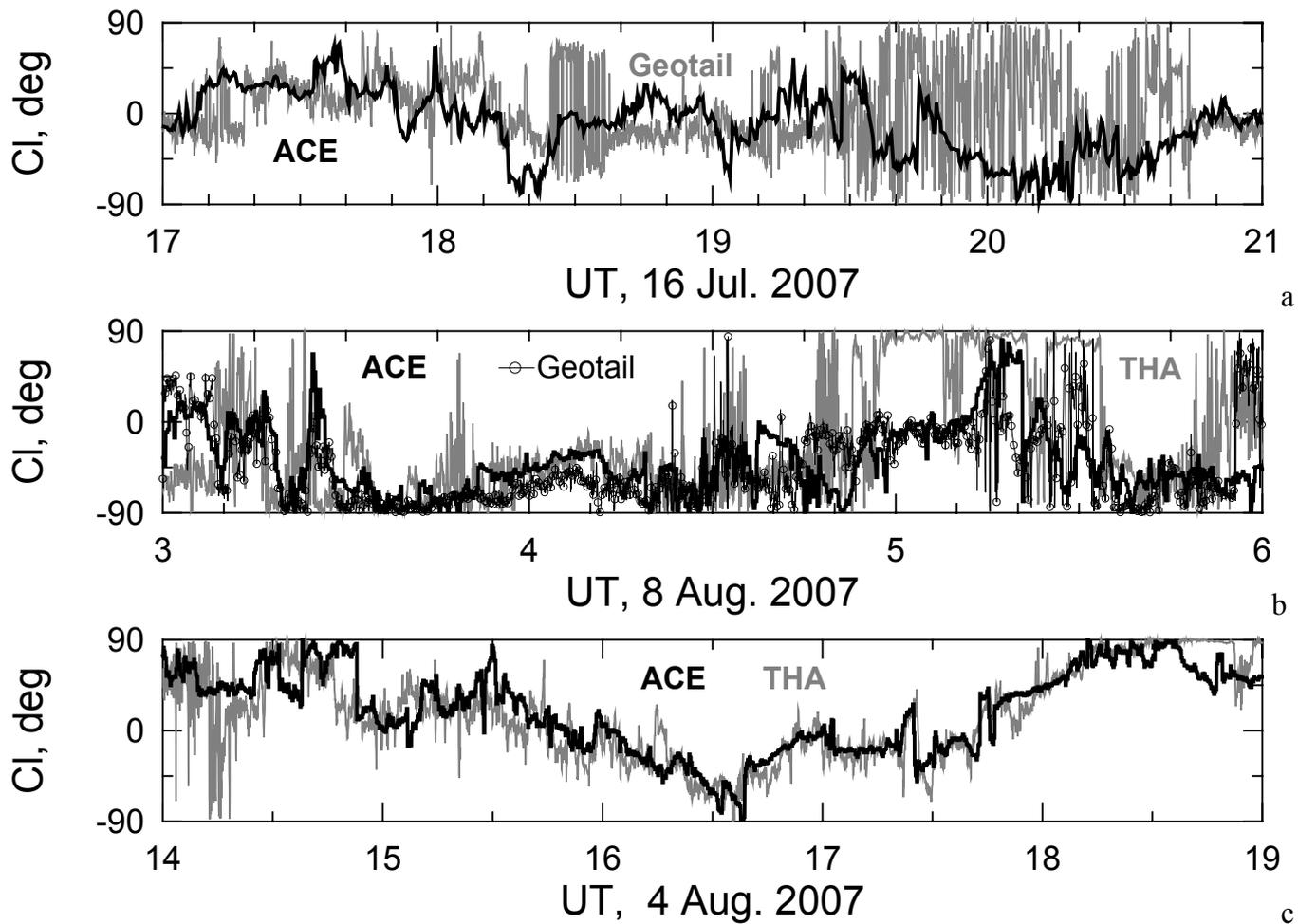

Fig.5

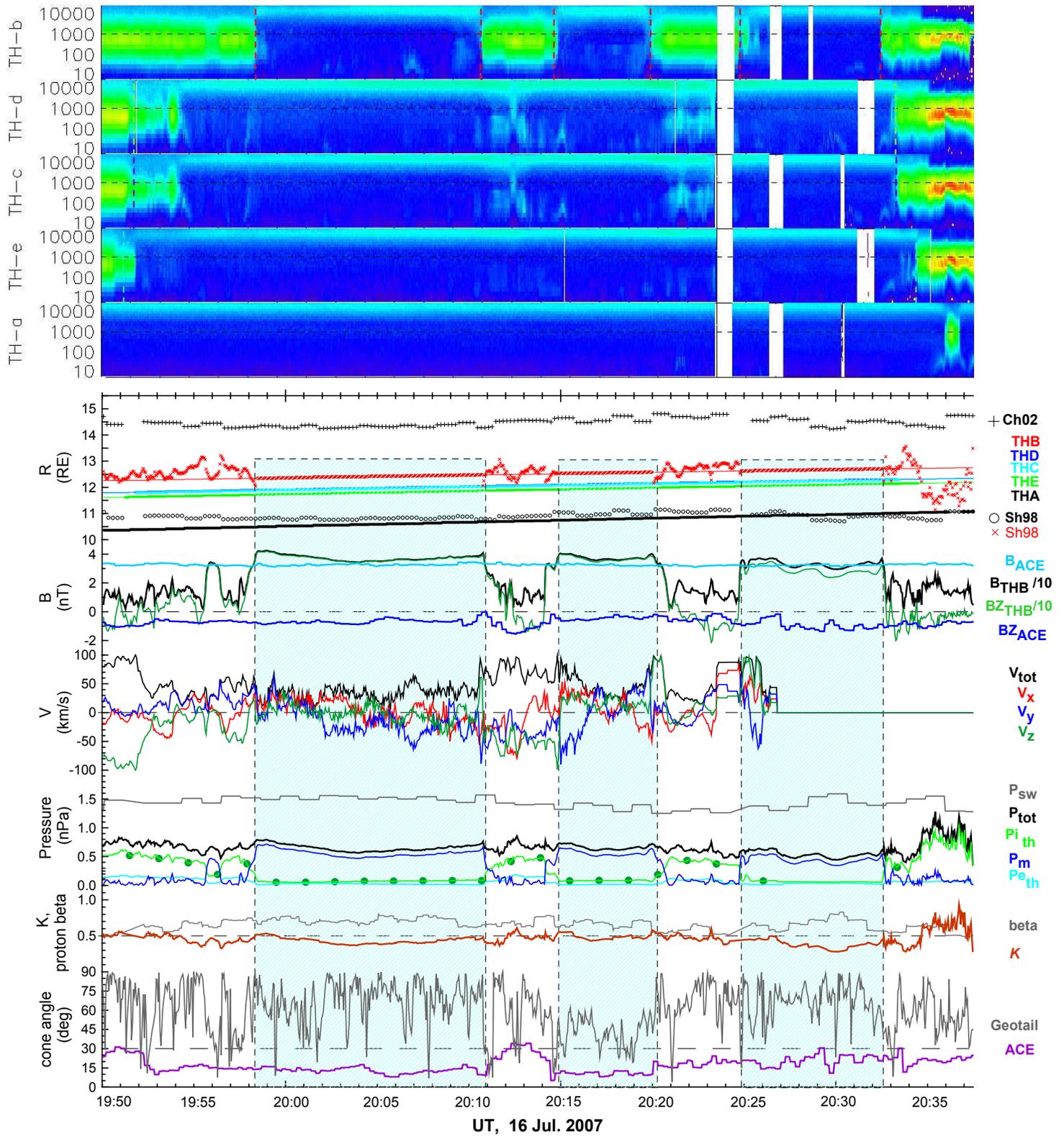

Fig.6

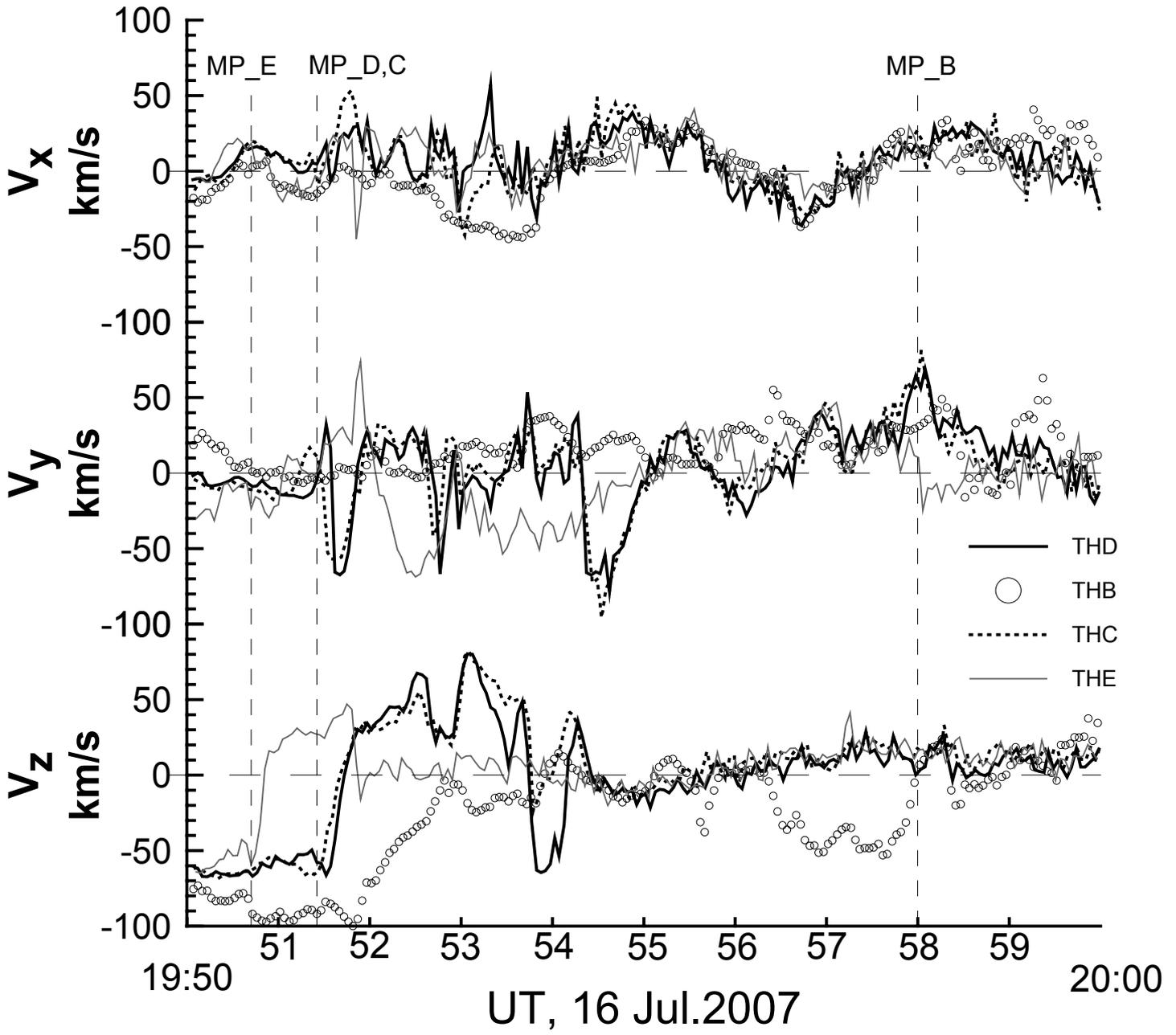

Fig7

Fig.8

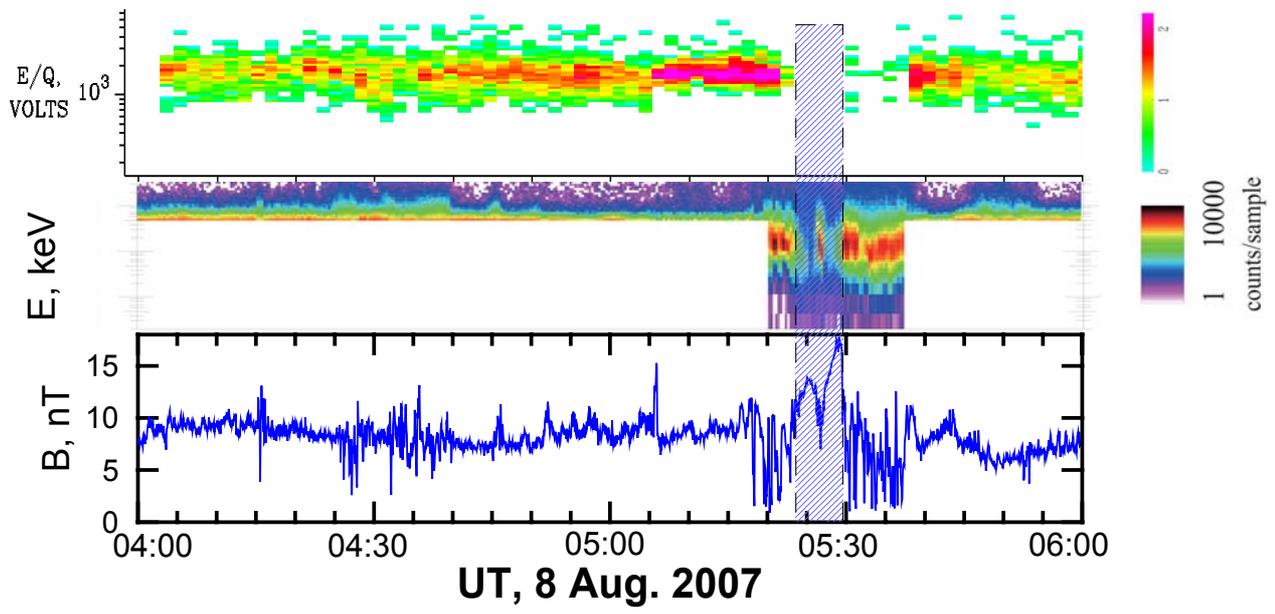

Fig.9

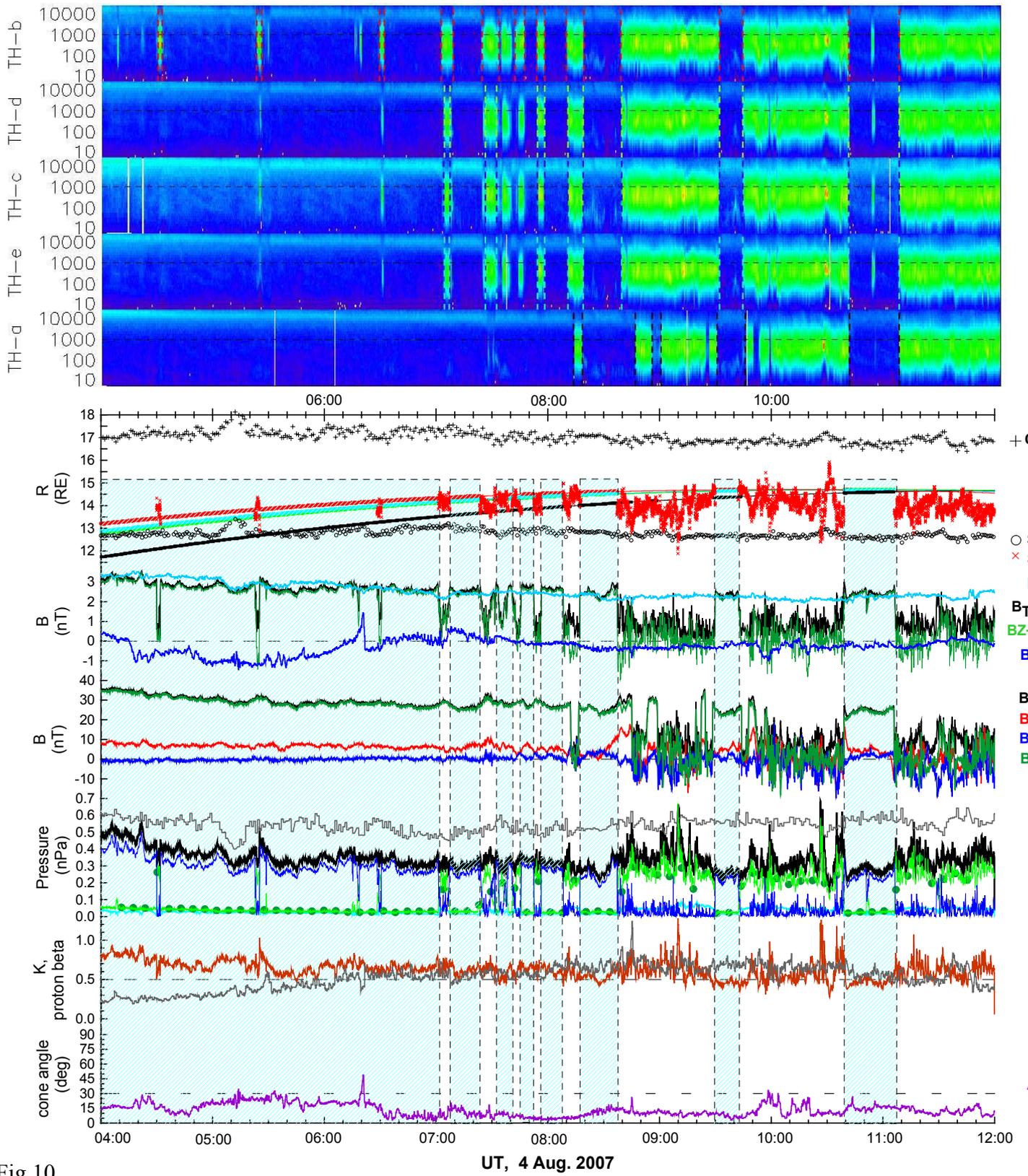

Fig 10

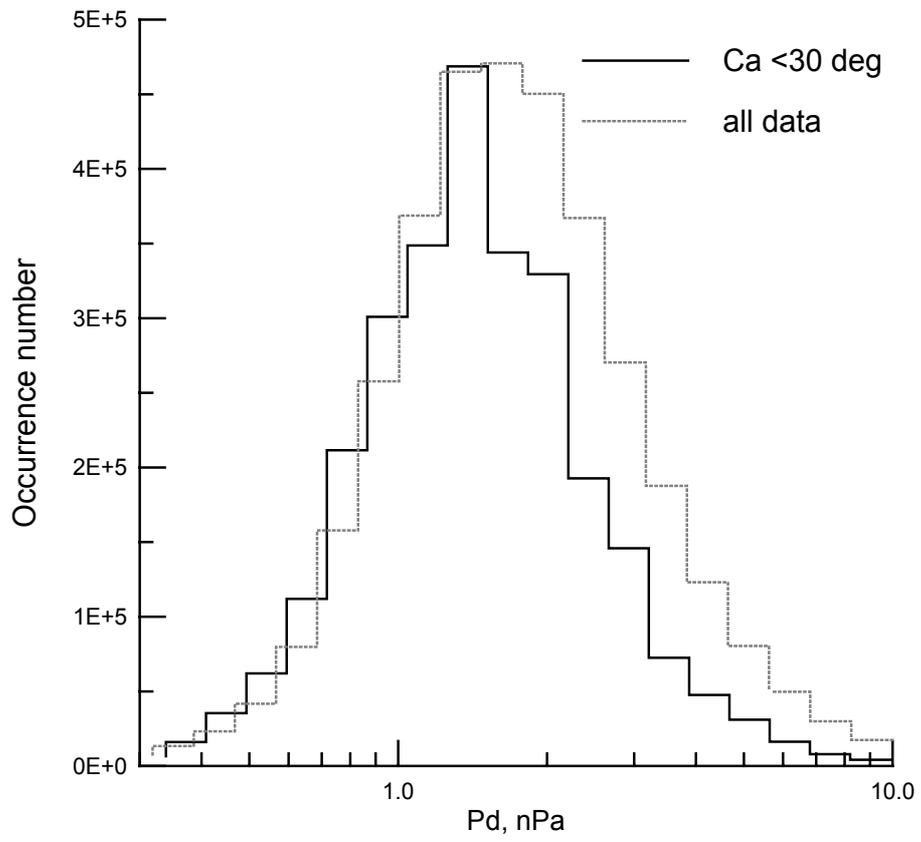

Fig. 11.

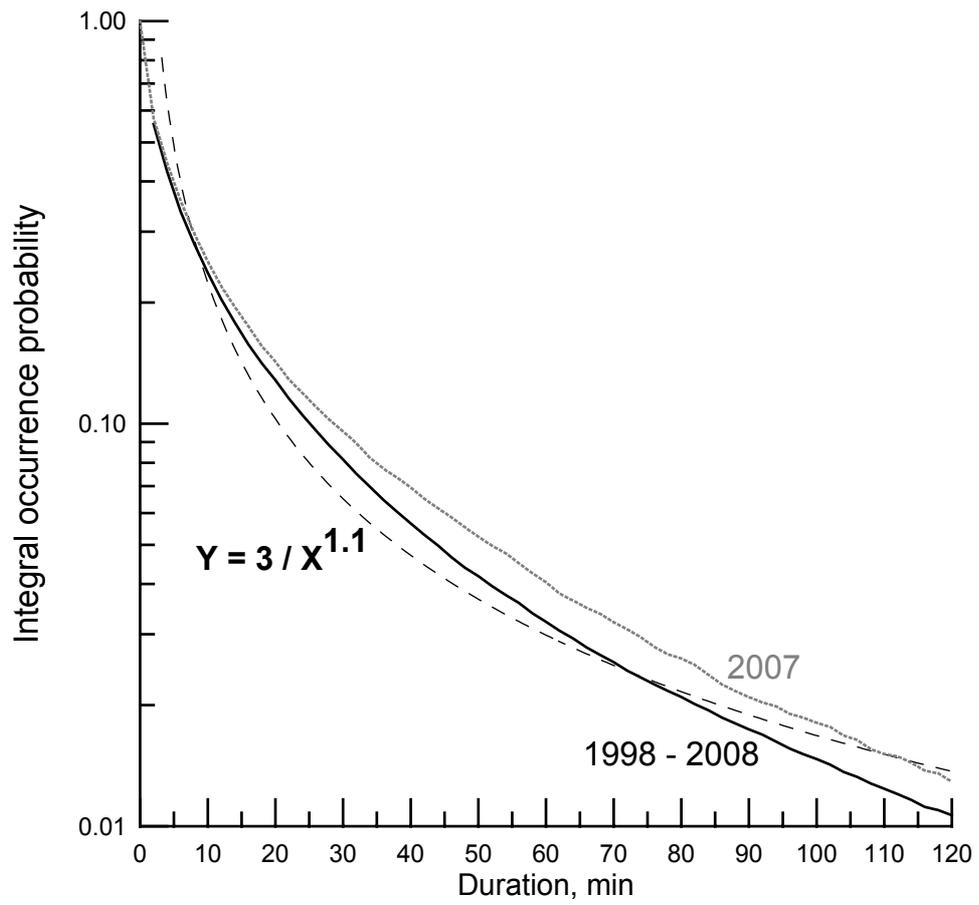

Fig. 12.

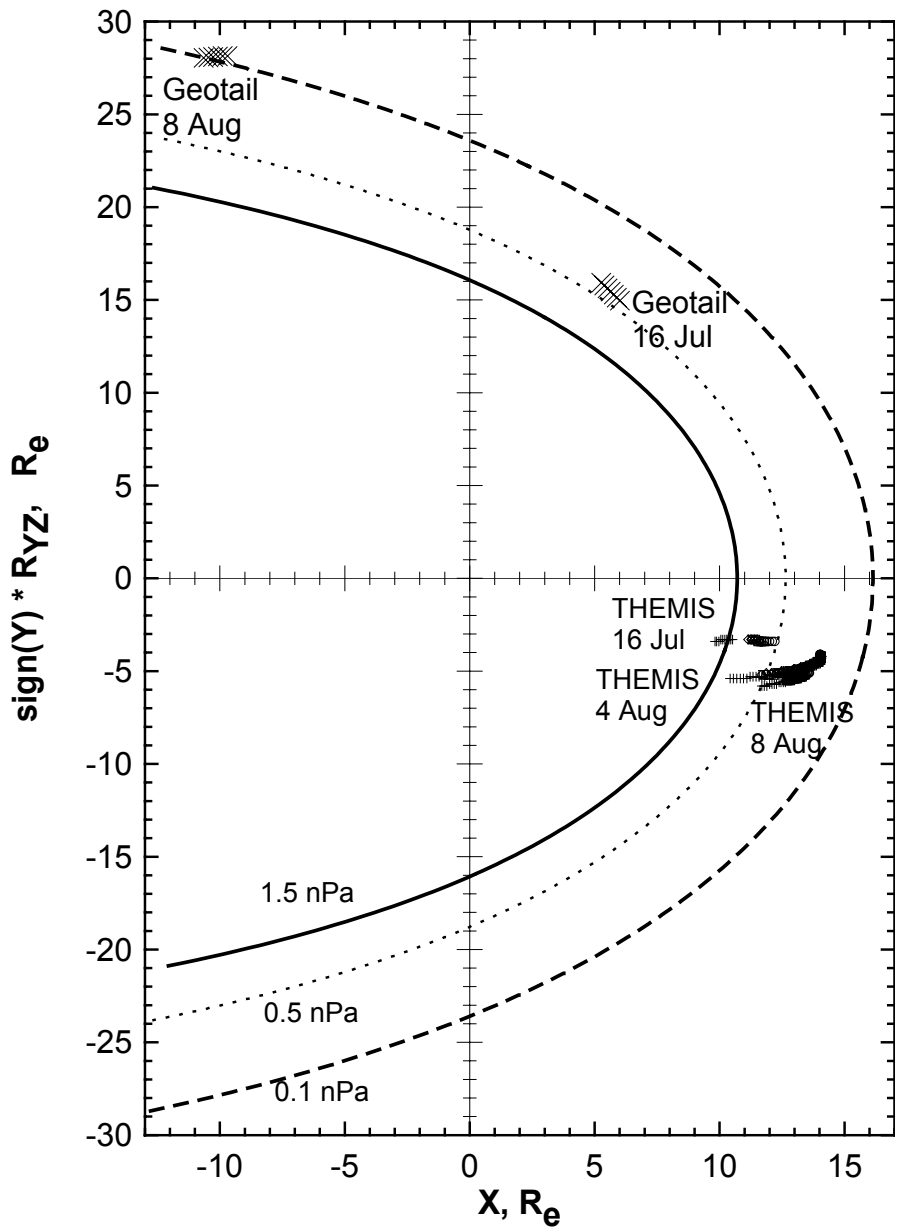

Fig.13